\documentclass[11pt,a4paper]{article}
\usepackage{graphicx}
\usepackage{amsmath}
\usepackage{cite}
\usepackage{verbatim} 
\usepackage[usenames]{color}
\usepackage{url}
\usepackage{lineno}[displaymath, mathlines]

\newcommand{\bEq}{\begin{equation}}
\newcommand{\eEq}{\end{equation}}
\newcommand{\bEQ}[1]{\begin{equation} \begin{array}{#1}}
\newcommand{\eEQ}{\end{array} \end{equation}}

\usepackage[vlined]{algorithm2e}

\newcounter{CorrectThis}

\begin{document}

\pagestyle{empty}

\null

\vfill

\begin{center}

{\Large  
{\bf On modeling global grain boundary energy functions}
}\\

\vskip 1.0cm
A. Morawiec
\vskip 0.5cm 
{Institute of Metallurgy and Materials Science, 
Polish Academy of Sciences, \\ Krak{\'o}w, Poland.
}
\\
E-mail: nmmorawi@cyf-kr.edu.pl \\
Tel.: ++48--122952854, \ \ \  Fax: ++48--122952804 \\

 \end{center}

\vfill


\noindent
{\bf Abstract}
\\
Grain boundaries affect properties of polycrystalline materials. 
The influence of a boundary is largely determined by its energy.
Grain boundary energy is often portrayed as a function 
of macroscopic boundary parameters  
representing grain misorientation and boundary plane inclination.
In grain boundary simulation and modeling, 
many studies neglect structural multiplicity of boundaries, i.e.,
existence of metastable states, and 
focus on minimum energy.
The minimum energy function restricted to 
constant misorientation 
should satisfy Herring's condition for interface stability.
This requirement has been ignored in recent works on 
grain boundary energy functions.
Example violations of the stability condition are shown. 
Moreover, a simple and natural procedure for constructing a continuous function 
satisfying the condition is described. 
Cusps in the energy as a function of boundary plane inclinations 
arise because of the imposition 
of the stability condition, and 
their locations and shapes result from properties of input data.
An example of applying the procedure to simulated data is presented.

\vskip 0.5cm

\\

\vskip 0.2cm


\newpage

\pagestyle{plain}

\noindent
\section{Introduction}

Knowing grain boundary properties is important for modeling microstructures of polycrystalline materials.
One of these properties is the boundary free energy. 
The practically feasible identification of boundary types  
is based on macroscopic boundary parameters 
(representing grain misorientations 
and boundary planes), and an energy function covering this 
parameter space is sought.
Attempts are being made to determine such functions experimentally 
(e.g., \cite{saylor2002distribution,li2009relative,beladi2014five,shen2019determining}), 
but obtaining results with acceptable resolution and reliability is difficult.
Another approach is to use 
atomic-scale simulations to compute energies of some pivotal boundaries 
and then use approximation methods to estimate energies of all other boundaries. 
It is well know from experiments and simulations
that boundaries have multiple metastable structures.
However, many studies on grain boundary 
simulation and modeling ignore structural multiplicity and deal with minimum energies.
Recent years have seen a proliferation of publications on
constructing global minimum energy functions 
based on data from simulations; see, e.g., 
\cite{kim2011identification,
	restrepo2014using,
	bulatov2014grain,	
	dette2017efficient,	
	sarochawikasit2021grain,
	zhang2022molecular,
	chirayutthanasak2022anisotropic,	
	chirayutthanasak2024universal,
	baird2021five,
	homer2022examination}.
Such functions have been used for polycrystal modeling, e.g.,  
for simulation of grain growth under anisotropic conditions 
\cite{kim2014phase,hallberg2019modeling,nino2023influence,nino2024evolution}.

When grain misorientation is constant, 
the formal description of
grain boundary energy as a function of boundary plane 
is analogous to that of surface energy.
The minimum energy functions satisfy the condition
for interface stability first precisely
formulated by Herring \cite{Herring_1951}.
This fact is well known,
but has somehow been ignored in works on 
global grain boundary energy functions.
Noncompliance with  Herring's condition is found in simulation data. 
In modeling of the energy function, 
noncompliance with the condition is
also caused by an inappropriate model of energy cusps.
For the consistency of boundary analysis with theory, 
it is important to recognize the nature of 
these contraventions and indicate ways to avoid them.

This paper recalls Herring's interface stability condition and 
shows examples of its violation.
Moreover, a simple procedure for constructing 
a continuous function that satisfies the condition is described.
The first step of the procedure is to obtain a  
function that interpolates between data points.
The interpolating function is then modified by imposing 
Herring's condition on function's restrictions 
to constant misorientations.
This may lead to cusps 
in energy as a function of boundary plane 
even if the interpolating function does not have them.
The procedure does not require entering cusp locations 
and there is no need for a cusp model.
For illustration, 
results of application of the procedure 
to simulated data are presented.
Examples below concern a single metal
with cubic $m\overline{3}m$ symmetry, but key aspects 
of the paper apply to a broader class of materials.

\section{Energy as a function of boundary plane inclination \label{section_2}}

The grain misorientation can be represented by a special orthogonal matrix, say $M$,
and the boundary plane by a unit vector, say $\mathbf{n}$, normal to the plane 
with components given in the reference frame linked to the first crystal.
The grain boundary free energy per unit area $\Gamma$ is considered to be a 
function of $M$ and  $\mathbf{n}$. 
The function is assumed to be continuous, non-negative and bounded from above, with 
singularities where the derivative of the function does not exist.
The singularities are usually referred to as energy cusps 
when shown in sections through $\Gamma$.
It is also assumed that $\Gamma$ decreases 
with decreasing misorientation angle
when the latter is smaller than a certain limiting value, 
and $\Gamma$ equals zero at zero misorientation angle.

Let the grain misorientation be fixed at $M$,
and let the variable vector $\mathbf{n}$ cover the sphere. 
The focus below will be on the function $\gamma$  
defined as 
$\gamma(\mathbf{n}) = \Gamma(M,\mathbf{n})$. 
Physically, the case with fixed misorientation can be viewed as a  
crystal embedded inside another crystal. 
The configuration  is similar to that considered 
with respect to the equilibrium shape of a crystal immersed in another medium.
Knowing $\gamma$, one can draw and analyze 
its spherical plot, i.e., the Wulff plot.

\begin{figure}
	\begin{picture}(300,580)(0,0)
		\put(25,410){\resizebox{6.2 cm}{!}{\includegraphics{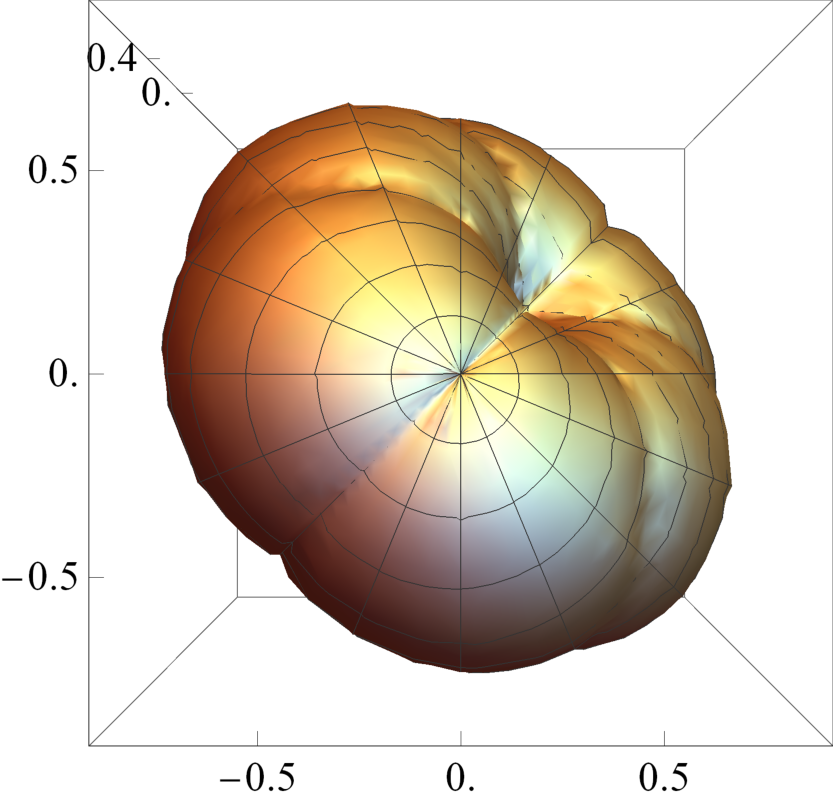}}}
		\put(32,185){\resizebox{5.8 cm}{!}{\includegraphics{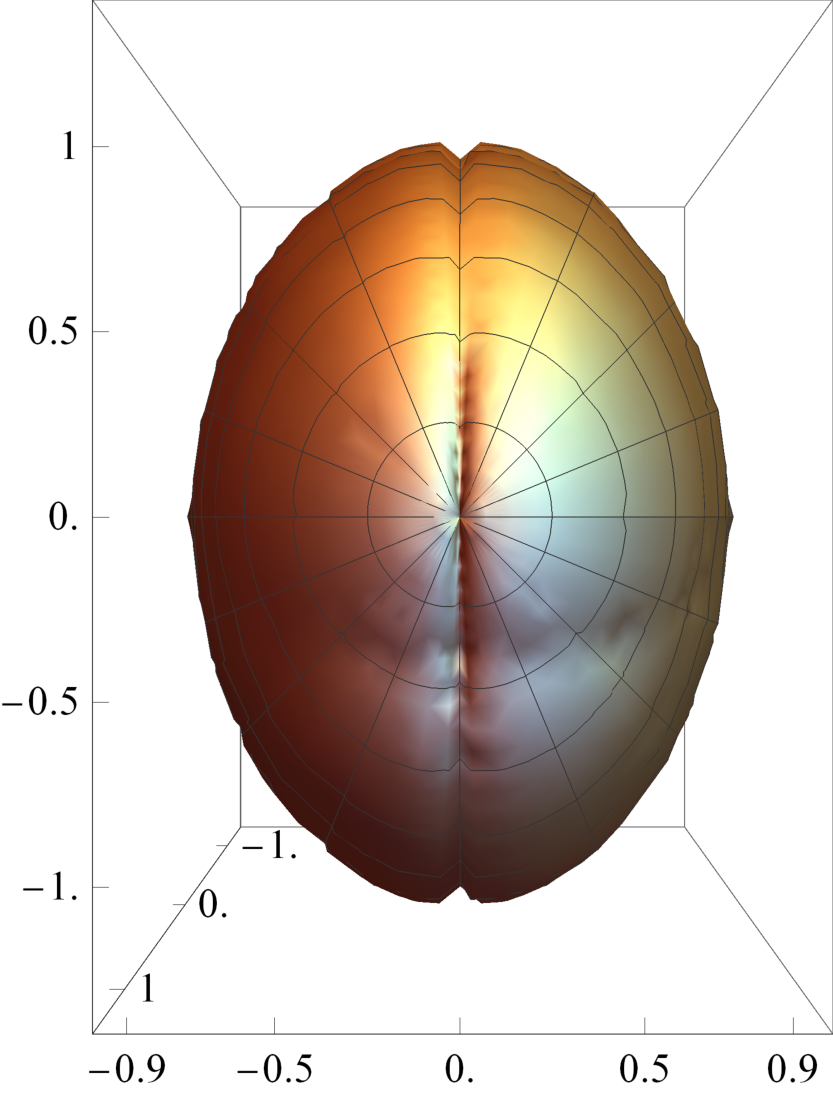}}}
		\put(25,0){\resizebox{6.2 cm}{!}{\includegraphics{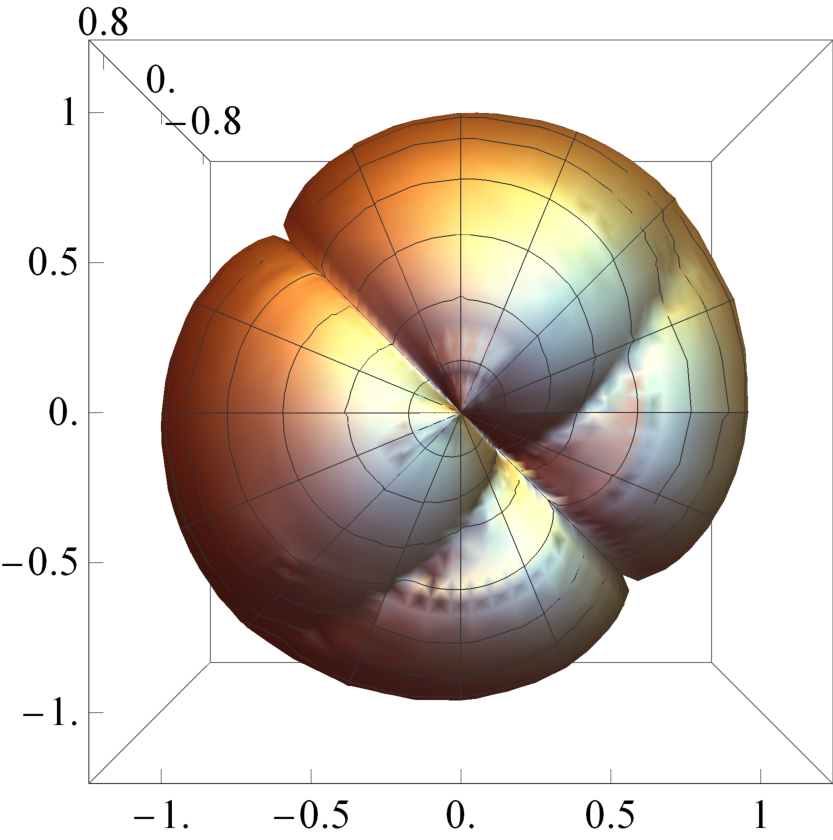}}}
		
		\put(240,400){\resizebox{5.7 cm}{!}{\includegraphics{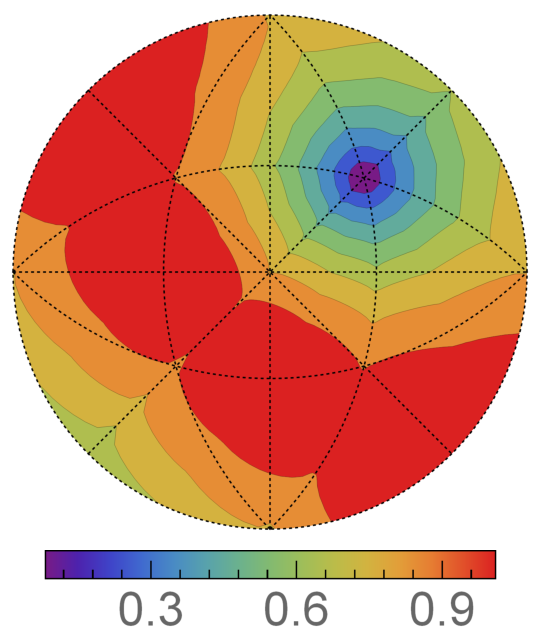}}}
		\put(240,200){\resizebox{5.7 cm}{!}{\includegraphics{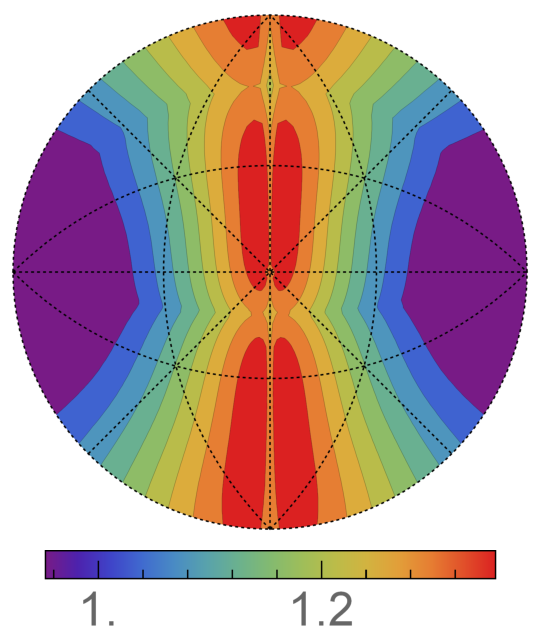}}}
		\put(240,0){\resizebox{5.7 cm}{!}{\includegraphics{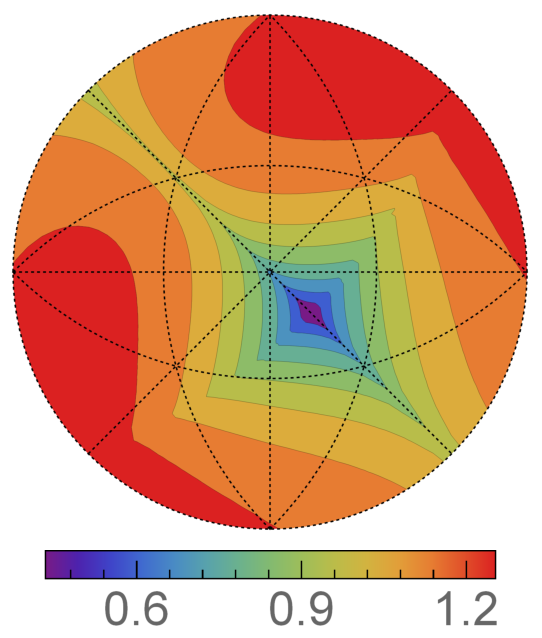}}}
		
		\put(10,575){\textit{a}}
		\put(10,385){\textit{b}}
		\put(10,175){\textit{c}}
	\end{picture}
	\vskip 0.0cm
	\caption{Wulff plots and their stereographic projections for Ni and the misorientations
		$\Sigma 3$, 
		i.e., $60^{\circ}$ rotation about $[1\,1\,1]$
		 (\textit{a}),
		$\Sigma 5$, 
		i.e., $\arccos(4/5) \approx 36.87^{\circ}$ rotation about $[1\,0\,0]$  
		(\textit{b}) and
		$\Sigma 11$, 
		i.e., $\arccos(7/11) \approx 50.48^{\circ}$ rotation about $[1\,1\,0]$    
		(\textit{c})
		based on the energy function of Bulatov et al. \cite{bulatov2014grain}.
		The stereographic projections shown here and in  
		Fig.~\ref{Fig_resulting_energy_function}
		were drawn using the program of G{\l}owi{\'n}ski 
		\cite{Glowinski_2015}.
		The energy unit is $\mbox{J/m}^2$. 
	}
	\label{Fig_example_Wulff_plots}
\end{figure}
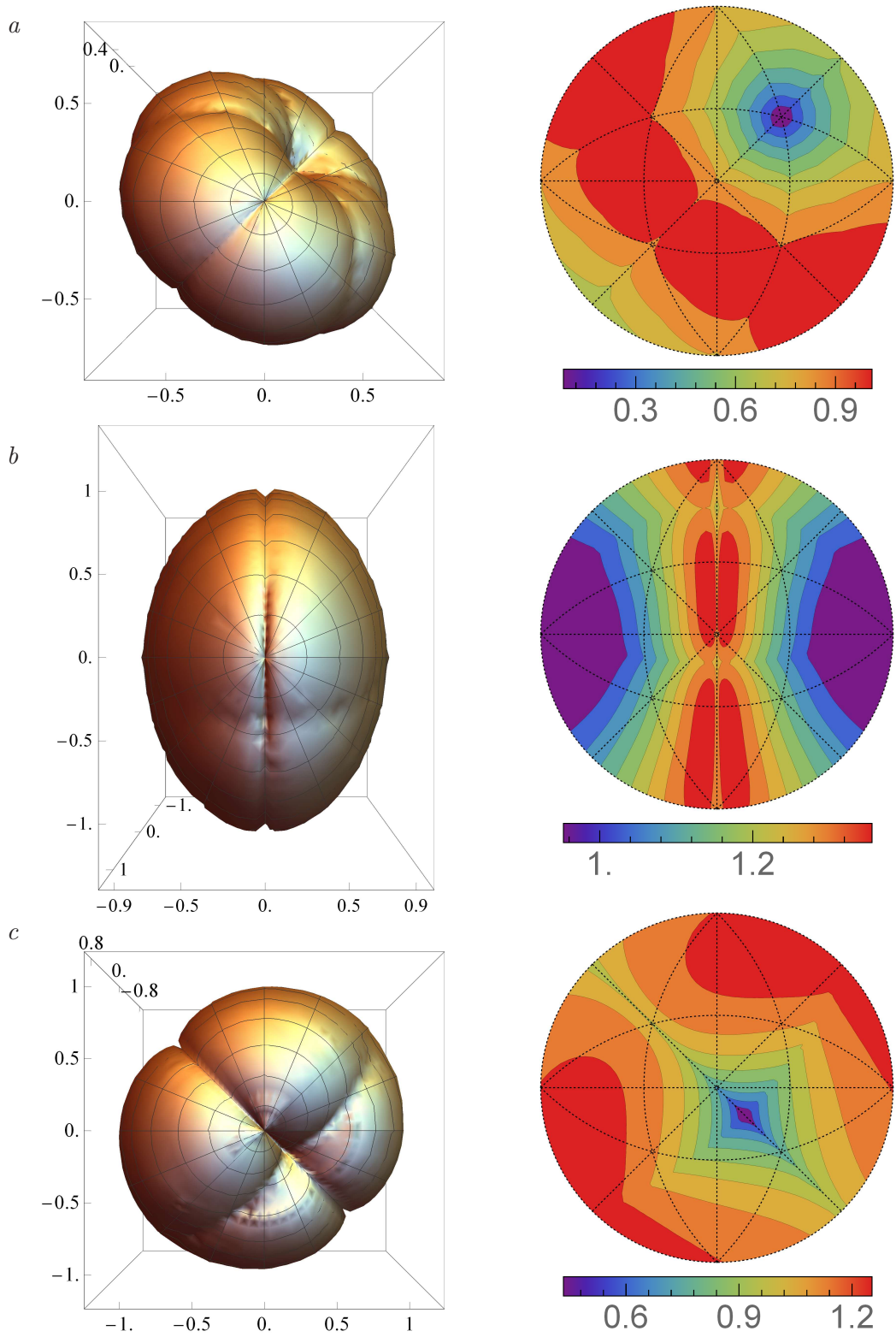

Example Wulff plots 
are shown in Fig.~\ref{Fig_example_Wulff_plots}.
These are sections through the closed-form  energy function described in the 
paper of Bulatov et al. \cite{bulatov2014grain}.
The function is based on atomic-scale simulations of Olmsted et al. \cite{olmsted2009survey} and
a-priori selected locations of 
cusps.
The function is constructed by what is called `hierarchical interpolation', 
but is actually a fitting, 
and the angular dependence of energy in the presence of a cusp
follows the model of Wolf \cite{Wolf_1989}. 
Fig.~\ref{Fig_example_Wulff_plots} and all other examples below are based on 
data for Ni taken from \cite{olmsted2009survey} and \cite{bulatov2014grain}.

\subsection{H-convexity of energy function}

When energy functions are constructed by approximating simulated data,
atomistic and continuum models are intertwined.
In the continuum model, the boundary energy is understood to be 
the excess free energy associated with the presence of the boundary 
per unit area. 
The question arises about the nature of the energy obtained in simulations.
Costs of computing grain boundary energy from first principles
is currently too high for any extensive coverage of the five-dimensional parameter space.  
Therefore, simplified computational schemes are used. 
Most estimations of grain boundary energy rely on 
athermal molecular dynamics simulations
at $0$K.
Typical determination of the boundary energy 
involves multiple (macroscopically equivalent) initial atomic configurations
\cite{olmsted2009survey,
	sarochawikasit2021grain,
	zhang2022molecular,
	chirayutthanasak2022anisotropic,	
	chirayutthanasak2024universal}. 
For each initial configuration, the simulation software 
searches for the configuration with minimum energy. 
This minimum energy usually varies with the initial configuration, 
meaning that energy minima are local and boundary states are metastable. 
In the end, the lowest of the energies obtained from all tested initial configurations 
is taken as the actual energy of the boundary under consideration.
Thus, in such simulations,  
the search is for `global minima' of the energy.  
Such minima correspond to equilibrium states.
The equilibrium states are rarely achieved in real materials 
and are difficult to reach in simulations, but their energies are unique,
while those of metastable states can spread over wide ranges.

It follows from the above that if a given 
value of boundary energy results from simulation 
aimed at finding minimum-energy configurations,
it should not be possible to assign any lower energy to this boundary.
However, this rule is not always met.
This can be demonstrated using standard Herring's \cite{Herring_1951} analysis 
of Wulff plots or its version based on 
$1/\gamma$-plots \cite{Frank_1963}. 
The short formulation of the second approach is that for 
the boundary normal to $\mathbf{n}$ to be in equilibrium, 
the point $\mathbf{n}/\gamma(\mathbf{n})$ 
must lie on the convex hull of the $1/\gamma$-plot.
A function on the unit sphere satisfying the above 
(Herring) condition will be called h-convex, and 
a pentavariate function $\Gamma$
will be referred to as h-convex if all
its restrictions $\gamma$ for misorientations held constant are h-convex.

\begin{figure}
	\begin{picture}(300,230)(0,0)
		\put(40,0){\resizebox{8.0 cm}{!}{\includegraphics{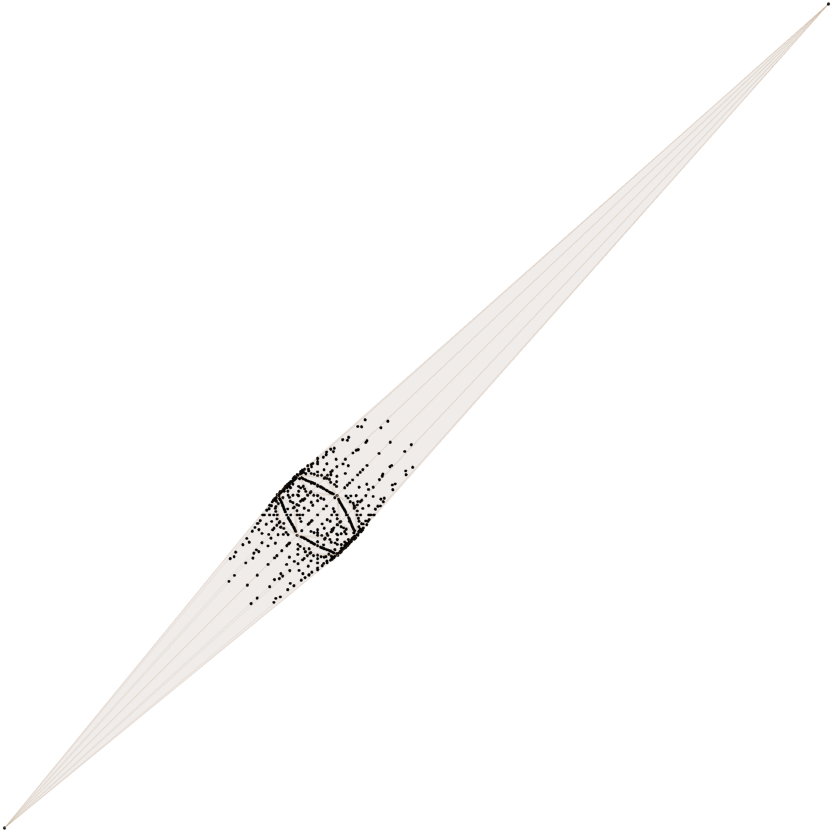}}}
		\put(0,110){\resizebox{4.0 cm}{!}{\includegraphics{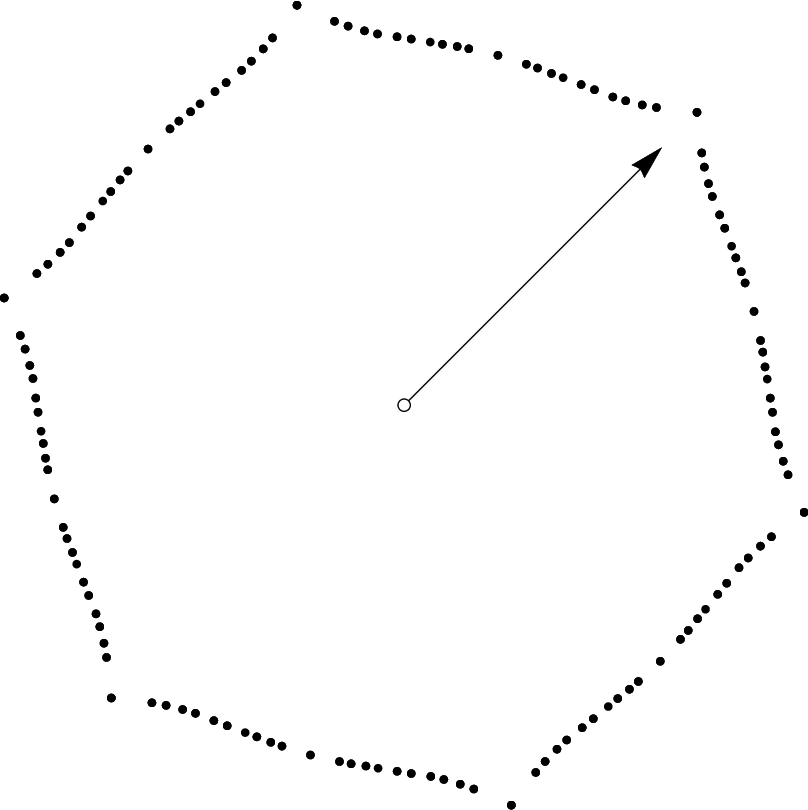}}}
		\put(260,0){\resizebox{6.0 cm}{!}{\includegraphics{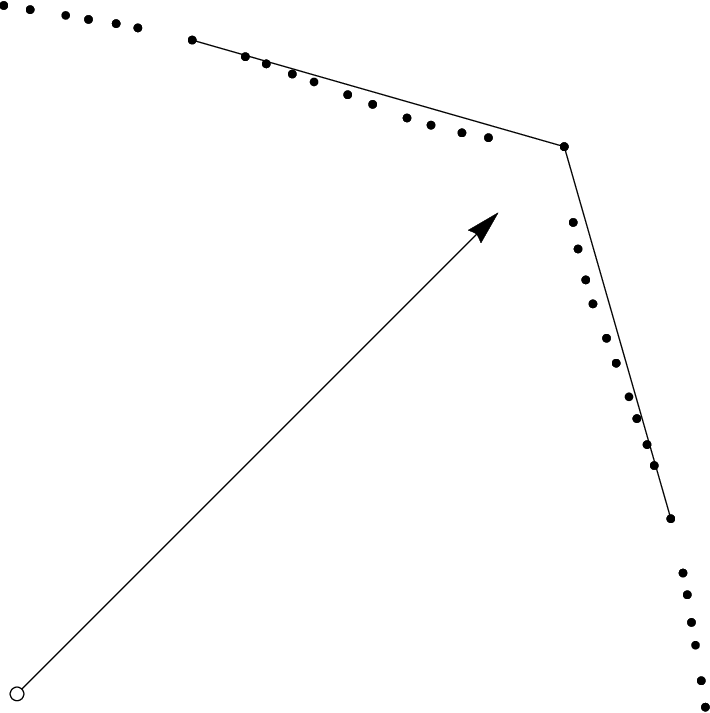}}}		
		\put(240,220){\textit{a}}
		\put(0,220){\textit{b}}
		\put(380,150){\textit{c}}
	\end{picture}
	\vskip 0.0cm
	\caption{The $1/\gamma$-plot based on data of Olmsted et al. \cite{olmsted2009survey}
		for Ni and the misorientation  $\Sigma 3$ (\textit{a}).
		To guide the eye, the convex hull of the point set is marked.
		The most distant vertices correspond to $\pm (1\, 1\, 1)$ planes.
		Section of the $1/\gamma$-plot perpendicular to $[ 1\, 1\, 1]$  
		and a part of it demonstrating that the plot is not convex are in 
		\textit{b} and \textit{c}, respectively.
		The vector shown in \textit{b} and \textit{c} 
		indicates  $( 1\, 1\, \overline{2})$ plane and its magnitude  is $1/(\mbox{J/m}^2)$. 
	}
	\label{Fig_err_Olm}
\end{figure}
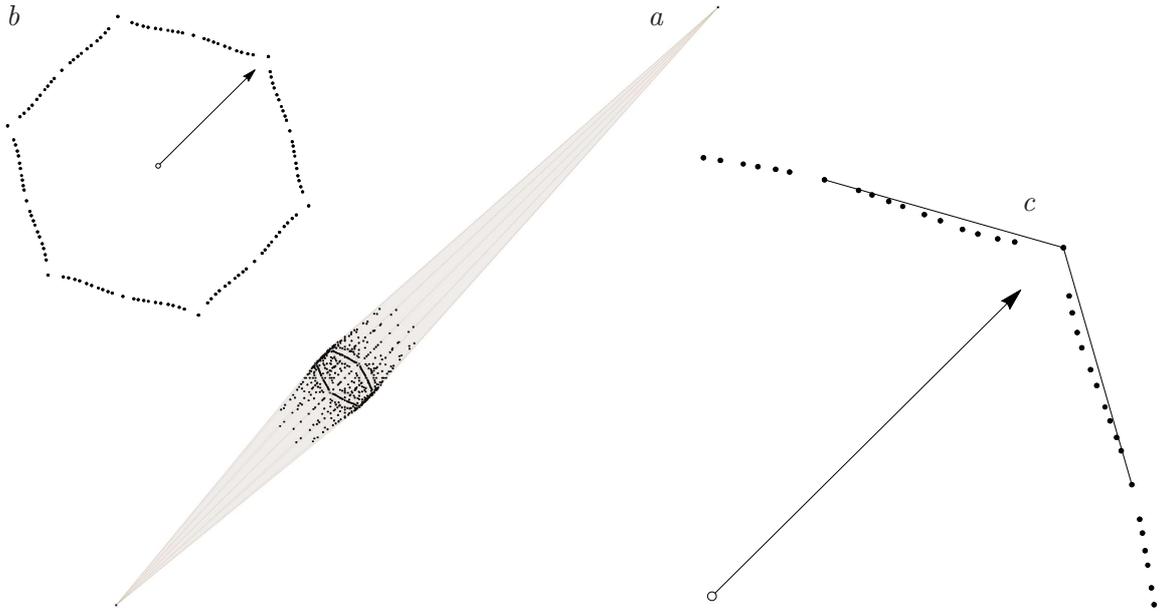

Example non-convexity of a $1/\gamma$-plot obtained based on the data set from \cite{olmsted2009survey}
is shown in Fig.~\ref{Fig_err_Olm}. 
Deviations from convexity are small, but quite common.
In the case of Ni and $M=\Sigma 3$, 
after taking into account symmetry \cite{Morawiec_1998,Morawiec_2009}, 
the number of distinct 
data points is 566, whereas the convex hull 
of the $1/\gamma$-plot has only 44 vertices.

The lack of h-convexity
means that 
some energies 
are too high.
In minimization of the crystal surface energy within the continuum model, 
such high energies imply missing surface orientations, i.e.,  
the corresponding regions of the Wulff plot are in a sense `passive', and 
"`passive' regions of the Wulff plot ... have no significance" \cite{rottman1984statistical}.
The same applies to grain boundary energies. 
The only characteristic energy of a boundary is
defined by the convex hull of the $1/\gamma$-plot. 
The passive regions of the Wulff plot 
correspond to Herring's `hill-and-valley' structures \cite{Herring_1951}.
In the case of boundaries, 
planes in passive regions are unstable and they facet.

On the other hand, simulated boundaries in passive regions
are stable against faceting. 
By using molecular statics at zero temperature, faceting is avoided, and the 
result of the computation is assumed to represent 
energy of the boundary plane selected at the outset.
In other words, the idea is that the simulation result is 
the lowest energy of non-faceted boundary 
with the plane matching the assumed macroscopic model.
However, by the nature of atomic simulations, 
the boundary is not flat, and it has a non-zero thickness,
i.e., it cannot match the plane exactly. 
The magnitudes of deviations of the true structure from the ideal 
flat zero-thickness macroscopic archetype vary from case to case,
and there are no definite limits on the allowed deviations.  
There is no 
clear-cut distinction between faceted and non-faceted forms of a boundary,
as boundary facets may vary in size from 
micro- and nano-scale facets observed in electron microscopes 
(e.g.  \cite{hsieh1989observations})
to `atomistic' facets of Brokman et al. \cite{brokman1981atomistic}
with facet recognition based on the observed arrangement 
of atoms within the structural unit of the boundary.
The results obtained from simulations incorporate the energies of 
facet edges and corners and their interactions.
The key point is that 
this energy depends on the density of the edges and corners. 
The density-dependent energy is not unique and it is not 
specific to the boundary defined by macroscopic parameters.
What is specific and unique is the limiting value 
to which the energy tends as the facet size increases.
This lower bound to energies the boundary may have
is the energy of the equilibrium configuration.

Similarly, continuous functions approximating 
simulated data sets containing global energy  minima for selected macroscopic boundary types
are expected to represent the energies 
of the equilibrium states.
The continuous function of Bulatov et al. \cite{bulatov2014grain} 
inherits non-covexities of the 
$1/\gamma$-plots from the data on which it is based, 
but this function has an additional reason for them to occur. 
It is the model of the energy cusp.

\section{Cusp model}

\begin{figure}
	\begin{picture}(300,470)(0,0)
		\put(120,310){\resizebox{6.0 cm}{!}{\includegraphics{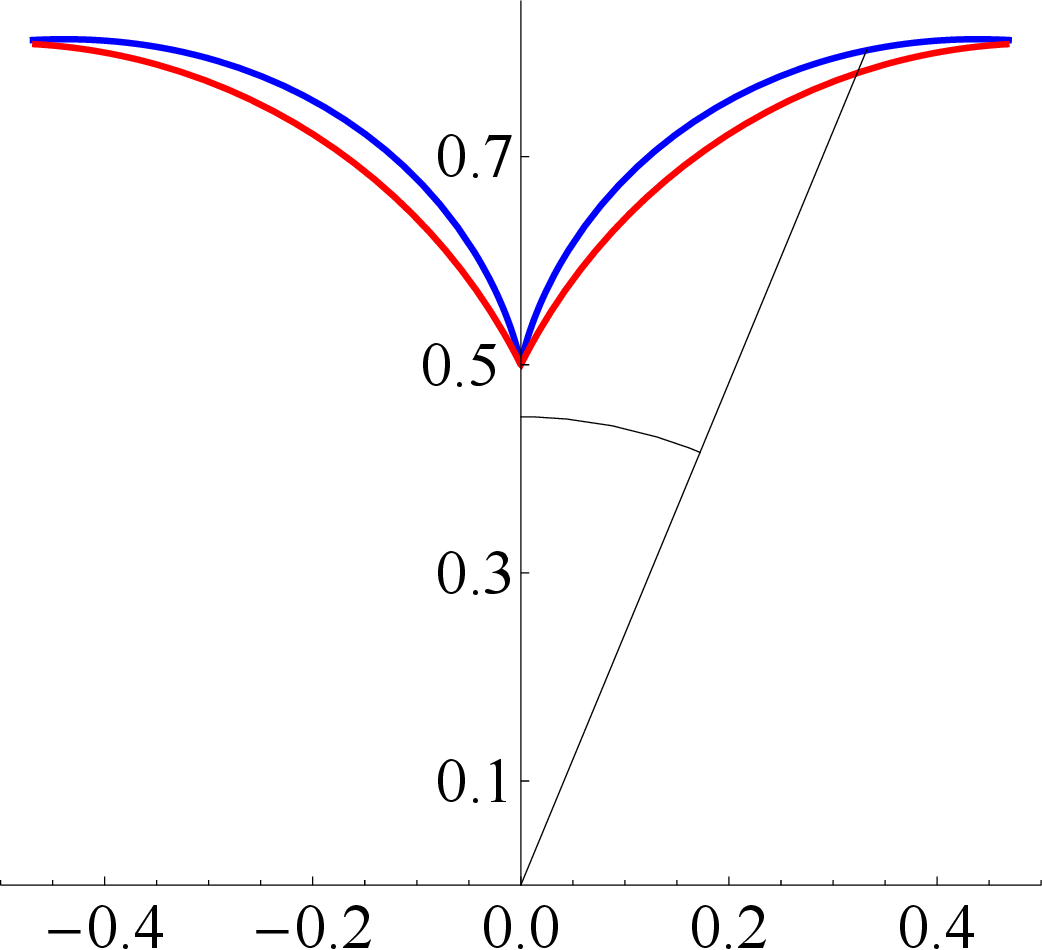}}}
		\put(120,0){\resizebox{6.0 cm}{!}{\includegraphics{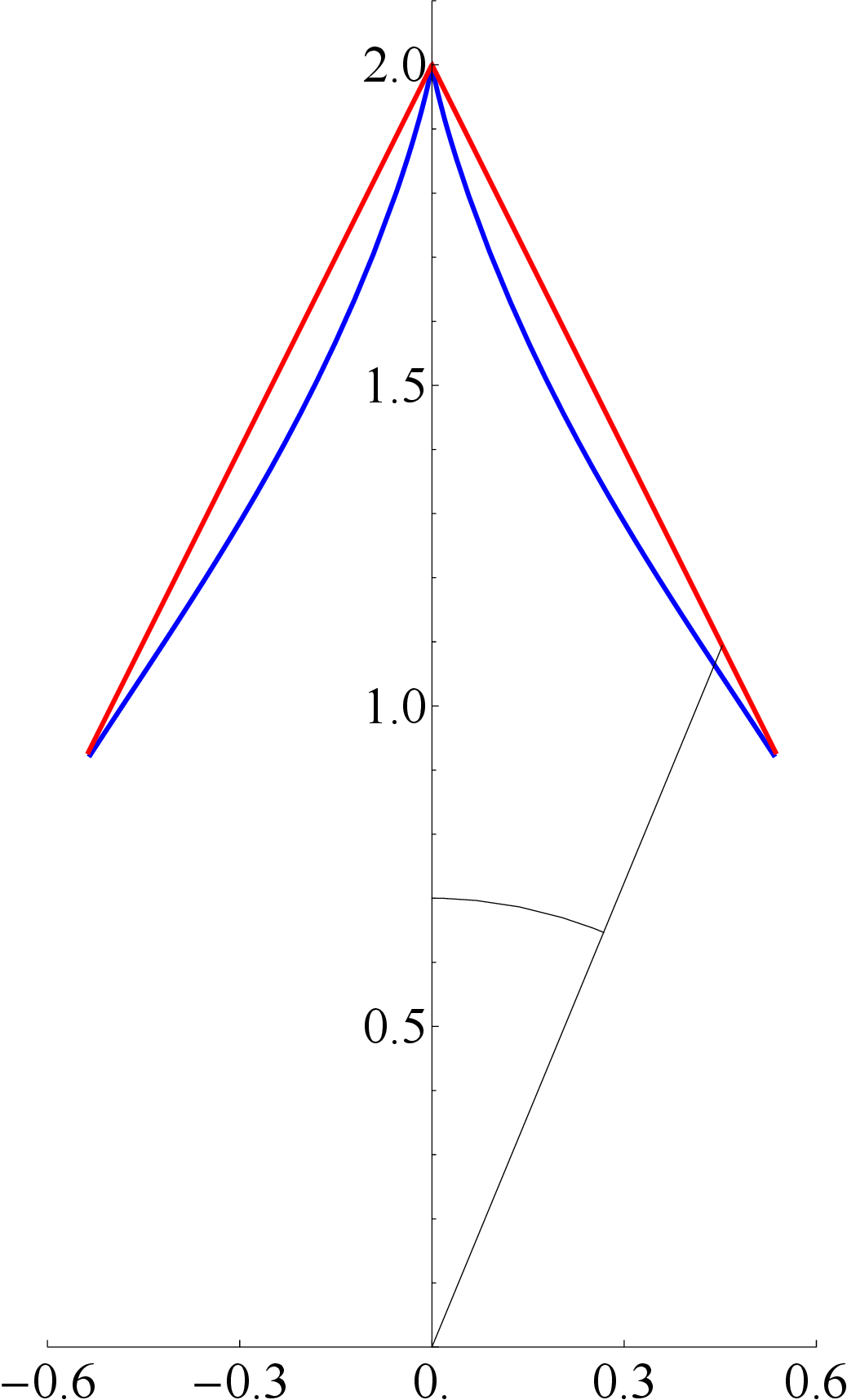}}}
		\put(216,70){$\phi$}
		\put(214,370){$\phi$}
		\put(110,460){\textit{a}}
		\put(110,270){\textit{b}}
		\put(224,456){$\gamma_{RSW}$}
		\put(235,431){$\gamma_c$}
		\put(244,198){$1/\gamma_c$}
		\put(216,156){$1/\gamma_{RSW}$}
	\end{picture}
	\vskip 0.0cm
	\caption{Parts of Wulff plots
		$\gamma_{RSW}(|\phi|) \times(\sin \phi,\cos \phi)$ (blue) 
		and
		$\gamma_c( |\phi |)\times(\sin \phi,\cos \phi)$ (red) 
		(\textit{a})
		and corresponding 
		$1/\gamma$-plots (\textit{b})
		for $a=0.5$, $b=0.65$, $g_0=0.5$, $g_{\pi/2}=1$, 
		and $\phi$ in the range form $-\pi/6$ to $\pi/6$.
		The angle between the branches of $\gamma_{RSW}$ at the 
		cusp is $0$, and the same applies to the branches of $1/\gamma_{RSW}$.
		The angle between the branches of $\gamma_c$ at the 
		cusp is $2 \arctan (g_0/g_{\pi/2})$, and the same angle is formed by the branches of $1/\gamma_c$.
	}
	\label{Fig_rsw}
\end{figure}

\begin{figure}
	\begin{picture}(300,550)(0,0)
		\put(142,440){\resizebox{5.2 cm}{!}{\includegraphics{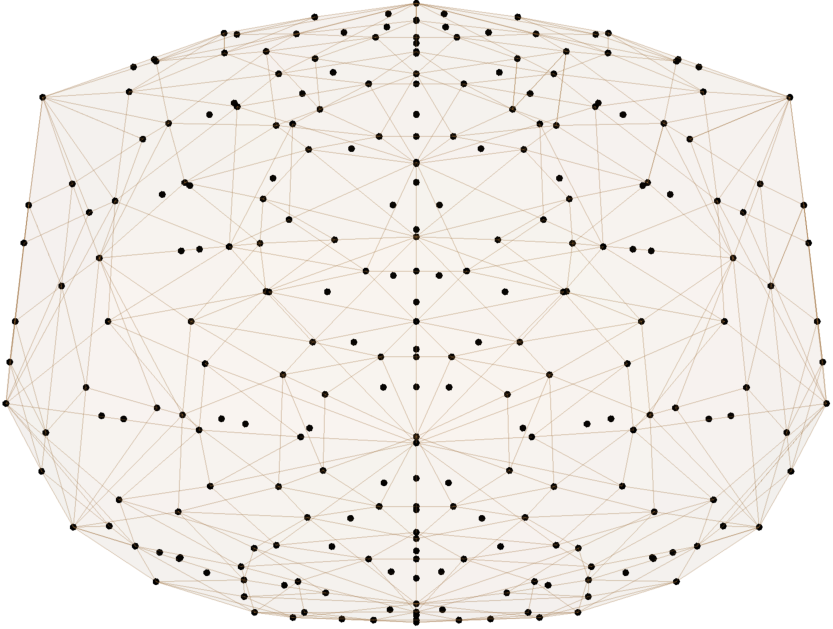}}}
		\put(90,250){\resizebox{8.0 cm}{!}{\includegraphics{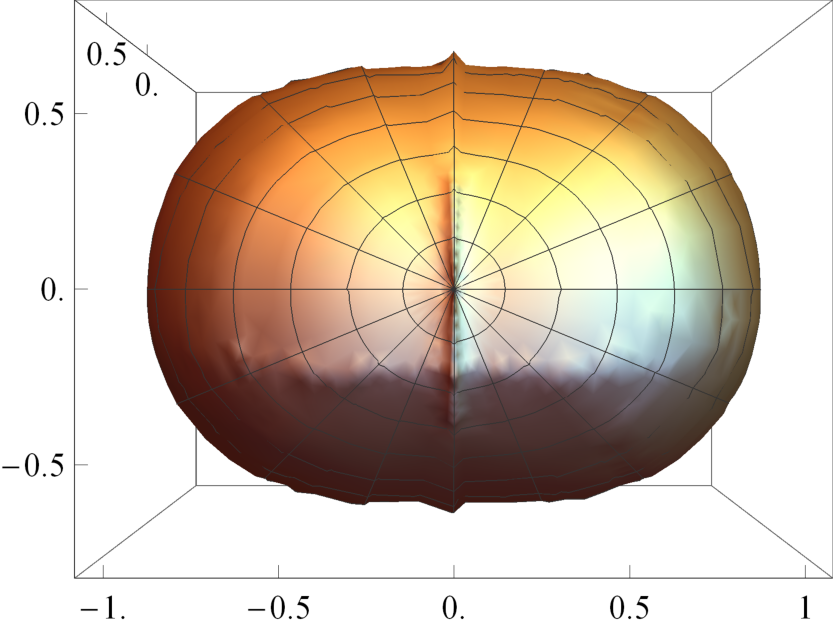}}}
		\put(128,0){\resizebox{6.0 cm}{!}{\includegraphics{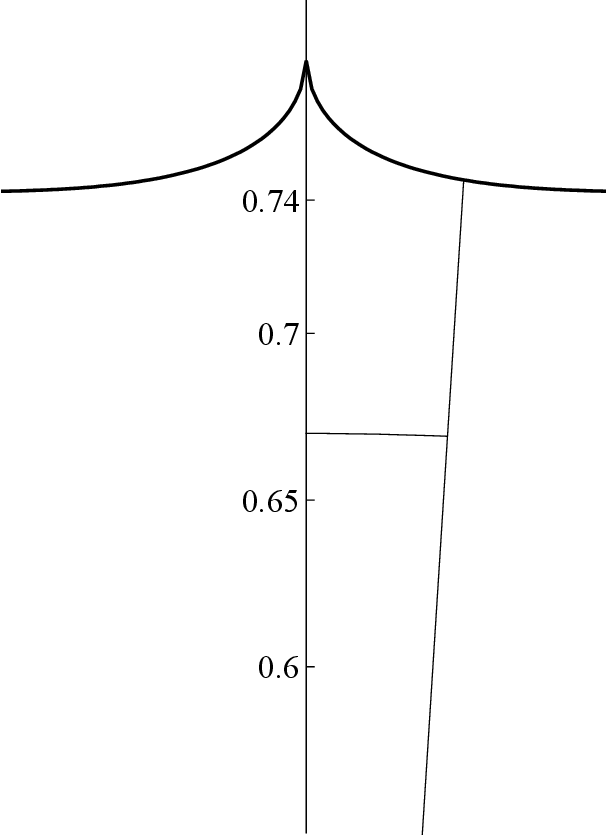}}}
		\put(70,540){\textit{a}}
		\put(70,410){\textit{b}}
		\put(70,210){\textit{c}}
		\put(230,80){$\phi$}
		\put(260,188){$1/\gamma$}
	\end{picture}
	\vskip 0.0cm
	\caption{
		(\textit{a}) The $1/\gamma$-plot based on data of Olmsted et al. \cite{olmsted2009survey}
		for Ni and the misorientation  $\Sigma 5$.		
		(\textit{b}) The corresponding $1/\gamma$-plot obtained from $\Gamma$ of 
		Bulatov et al. \cite{bulatov2014grain}. It has a ridge at directions 
		perpendicular to the misorientation axis $[1 \, 0 \, 0 ]$.
		(\textit{c}) Section through the plot shown in (\textit{b})  perpendicular to the ridge. It illustrates non-convexity of $1/\gamma$-plot. 
		The energy unit is $\mbox{J/m}^2$. 
	}
	\label{Fig_err_Bula}
\end{figure}
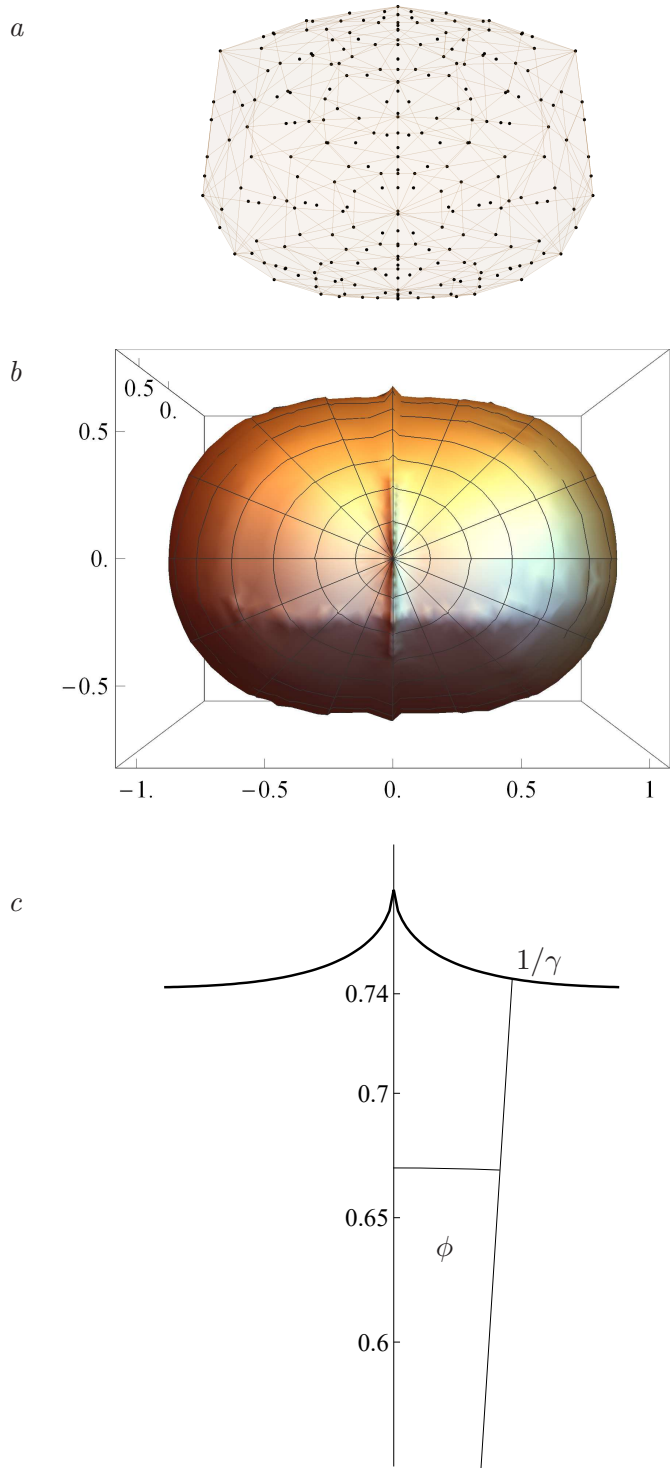

The shape of energy  near cusps is frequently 
modeled by the Read-Shockley function
$E_0 \,  \theta \left(A_1 -  \ln  \theta \right)$, where 
$\theta$ denotes the angle of misorientation from the position of the cusp, 
and $E_0$ and $A_1$ are parameters
\cite{Read_1950}.
Closely related is the Read-Shockley-Wolf (RSW) model.
In a slightly simplified form, 
with 
$a$, $b$ and $g_0$ denoting model parameters,
it is given by
$\gamma_{RSW}(\theta) = g_0 + b \sin \theta \, \left( 1 -  a \ln  \sin \theta \right)$
\cite{Wolf_1989}.\footnote{Assuming small $\theta$, Read and Shockley approximated $2\sin(\theta/2)$ by $\theta$ \cite{Read_1950}.
In Wolf's paper \cite{Wolf_1989}, $2\sin(\theta/2)$ is replaced by $\sin \theta$.
This gives zero slope at $\theta=\pi/2$  and simplifies the use of $\gamma_{RSW}$ over the range $0 \leq \theta \leq \pi/2$.}
In \cite{bulatov2014grain,dette2017efficient,sarochawikasit2021grain,zhang2022molecular,chirayutthanasak2022anisotropic,chirayutthanasak2024universal},
the use of the RSW model is extended beyond misorientation angles,
and it affects arbitrary cusps in $\Gamma$
including those with respect to inclination parameters, i.e., 
cusps in $\gamma$ functions at fixed misorientations.
It is easy to see, however,  that
$\gamma_{RSW}$ has an infinite slope at the cusp, and in effect, 
the $1/\gamma$-plot near an RSW-modeled cusp is not convex;
see Figs.~\ref{Fig_rsw} and \ref{Fig_err_Bula}.
Therefore, $\gamma_{RSW}$
is not suitable for modeling cusps in $\gamma$ functions.

The question is what is the formula for a cusp in $\gamma$
when it arises due to an edge of flat faces of the $1/\gamma$-plot.
It can be easily obtained by considering a two-dimensional case. 
Let $\mathbf{n}_c$ be the direction 
at which two faces of $1/\gamma$-plot meet, 
let the vector $\mathbf{n}$ vary in a fixed plane 
containing $\mathbf{n}_c$, and let
$\phi$ ($0 \leq \phi \leq \pi/2$) denote the angle between $\mathbf{n}_c$ and $\mathbf{n}$.
Then the energy function has the form
\begin{linenomath*}
\bEq
\gamma_c(\phi) = g_0 \cos \phi + g_{\pi/2} \sin \phi \ , 
\label{eq:cusp_shape}
\eEq
\end{linenomath*}
where $g_0 \ (> 0)$ is the energy at the cusp, and  
the derivative of $\gamma_c$ at the cusp is $g_{\pi/2}\ (> 0)$, i.e., 
the slope is finite (Fig.~\ref{Fig_rsw}).
With $0 \leq \phi \leq \pi/2$,
the curve 
$\gamma_c( \phi ) \, (\sin \phi,\cos \phi)$
(seen as a fragment of two-dimensional Wulff plot)
is a part of the circle 
centered at $(g_{\pi/2}, g_{0})/2$ and passing through the points 
$(0,0)$, $(0, g_{0})$ and $(g_{\pi/2},0)$; cf. \cite{Herring_1951}. 
By definition of $\gamma_c$, the curve 
$(\sin \phi,\cos \phi)/\gamma_c( \phi )$
(seen as a fragment of two-dimensional $1/\gamma$-plot)
is a segment of a straight line
through $(0,1/g_{0})$ and $(1/g_{\pi/2},0)$.

\section{H-convex energy model}

The essence of boundary energy modeling is to construct 
a global energy function based on discrete data.
The function, satisfying conditions listed in section \ref{section_2},
is expected to be not only close to the simulated data but also h-convex.

Let $\gamma_0$ be a function on the unit sphere. 
If $\gamma_0$ violates the condition
of h-convexity,
it can be modified so the condition is satisfied: 
each point inside the convex hull of the $1/\gamma_0$-plot 
is replaced by that on the convex hull.
In other words, 
given the function $\gamma_0$,
one must first obtain the $1/\gamma_0$-plot and determine
its convex hull,    
and the modified function is given by $\gamma(\mathbf{n})=1/q(\mathbf{n})$,
where $q(\mathbf{n})$ is the distance  
from the origin of the reference frame to the hull in the direction $\mathbf{n}$. 
This modification will be called h-convexification.
Note that to compute the value of the h-convex $\gamma$ 
for a single direction $\mathbf{n}$, 
it is necessary to know $\gamma_0$ over the entire sphere.
The important point is that the new h-convex function $\gamma$ may have 
singularities even if $\gamma_0$ is differentiable. 
Singularities will arise when 
the convex hull of the $1/\gamma_0$-plot
has vertices and/or edges;
the function $q$
is not differentiable at directions corresponding to such points, 
and this means that $\gamma$ is also singular at these directions.

In practice, it is convenient to h-convexify $\gamma_0$ using a discrete set of directions.
One needs to
generate a sufficient number $N_P$ of auxiliary vectors $\mathbf{n}_p$ 
on the unit sphere, 
get the set $P$ of points 
$\mathbf{n}_p/\gamma_0(\mathbf{n}_p)$ 
$(p=1,2,\ldots,N_P)$,
and determine its convex hull, i.e.,  
the convex polyhedron (seen as a union of convex polygons \cite{oRourke1998}) 
that encloses $P$
and whose vertices are in $P$.\footnote{
	The distance from the origin to a point on a face of 
	the polyhedron can be seen as a linear interpolation 
	over distances to the vertices of that face.
	H-convexification of the function $\gamma_0$ can be seen as \textit{linear} interpolation 
	over extreme points of the convex hull of 
	the  point set $P$.}
Natural candidates for the vectors $\mathbf{n}_p$
are normals to low-index lattice planes.  
Clearly, with this approach, presence of singularities in the resulting h-convex function
is certain, and cusps have shapes described by 
eq.(\ref{eq:cusp_shape}).

Let 
h-convexification of a pentavariate function $\Gamma_0$
defined on the entire space of macroscopic parameters
mean h-convexification of its restrictions to functions $\gamma_0$
for constant misorientations. 
A simple and natural method to construct a proper energy function $\Gamma$
is by h-convexifying a preliminary function 
that well approximates the discrete simulation data.
So the first step is to devise a function $\Gamma_0$ 
that will give an approximate energy at any point 
$\mathbf{B}$ based 
on known energies at points $\mathbf{B}^{\kappa}$ ($\kappa=1,2,\ldots,K$).
One may ask why introduce an approximation function $\Gamma_0$ 
instead of directly h-convexifying the discrete data set consisting 
of known energies at points $\mathbf{B}^{\kappa}$.
The reason is that for some misorientations the number of known energies is small, 
and for most (or precisely `for all except for finitely many') 
misorientations there are no data points at all.

One way to get $\Gamma_0$ is by interpolation. 
It is clear that 
in order to estimate $\Gamma$
at a single point, say $\mathbf{B}=(M, \mathbf{n})$,
one must first calculate the interpolated values $\Gamma_0$ 
for $N_P$ points $\mathbf{B}_p=(M,\mathbf{n}_p)$.
The number $N_P$ affects the computation time, 
but if it is large enough, 
it has no practical effect on the shape of $\Gamma$. 
The scheme for computing the value of h-convex $\Gamma$ 
for a given $\mathbf{B}$ is outlined as Algorithm I below.

H-convexification may cause singularities with respect to boundary plane inclination. 
There is a question if the function $\Gamma$ resulting from h-convexification 
of $\Gamma_0$ differentiable with respect to misorientation parameters
will also have this property.
There is no simple answer, as in 
cases other then 'twists' around normals to fixed planes, 
a change of misorientation implies a change in (at least one of) boundary planes.

It is worth making a digression on taking into account 
supplementary information on boundary energy.
A simple way to do this is by appending the simulated data with 
extra points carrying the information. 
It is natural to add the point with zero energy at the configuration 
$\mathbf{B}^0$
with zero misorientation angle.
Similarly, if there is little 
data for low-angle boundaries,
one may want to include a low-angle boundary model
by appending extra points with energies based on the model
to the initial 
data set. 
Merging datasets is a natural method for combining 
boundary data from different sources (e.g., simulation and experimental data).

\SetKwInput{KwInput}{Input}                
\SetKwInput{KwOutput}{Output}              

\begin{algorithm}
	
	\vskip 0.2cm
	\DontPrintSemicolon
	\SetAlgoRefName{I} 
	\KwInput{Parameters of the boundary $\mathbf{B}$ whose energy is to be calculated}

	\KwInput{List of $K$ boundaries $\mathbf{B}^{\kappa}$ and 
		corresponding energies $\Gamma_0^{\kappa}$}
	
	\KwInput{Subsidiary parameters $N_P$ and $u$}
	
	\BlankLine
	
	\tcc{Initialization}

	Decompose $\mathbf{B}$ into $(M,\mathbf{n})$
	
	Generate $N_P$ auxiliary directions $\mathbf{n}_p$

	\BlankLine
	
	\tcc{Interpolation between irregularly spaced data points}
	
	\tcc{For demonstration purposes, inverse distance weighting is used}

	Define the distance function $\widetilde{d}$

	\For{ $p=1,2,\ldots, N_P$}{
		
		$\mathbf{B}_p=(M,\mathbf{n}_p)$
		
		\For{ $\kappa=1,2,\ldots, K$}{
			
			$d_{\kappa}=\widetilde{d}(\mathbf{B}_p, \mathbf{B}^{\kappa})$
			
		}
		
		$\gamma_0(\mathbf{n}_p) =\sum_{\kappa} d_{\kappa}^{-u} \, \Gamma_0^{\kappa}/\sum_{\kappa} d_{\kappa}^{-u}$ \ \ \ \ \ \ \ \tcc{\  $=\Gamma_0(\mathbf{B}_p)$}
	}
	
	\BlankLine
	
	\tcc{Determination of the point set $P$}
	
	\For{ $p=1,2,\ldots, N_P$}{
		
		 $\mathbf{x}_p = \mathbf{n}_p/\gamma_0(\mathbf{n}_p)$		
	}
	
	\BlankLine
	
	\tcc{Determination of the convex hull of $P$}
	
	Call routine that determines the smallest convex polyhedron enclosing all 
	points $\mathbf{x}_p$	
	
	\BlankLine
	\tcc{Final step}	
	
	 $\mathbf{x}$ = the point of intersection of the 
	line $\lambda \mathbf{n}$ ($\lambda >0$)
	with the polyhedron	
	
	$q(\mathbf{n})=|\mathbf{x}|$
	
	$\gamma(\mathbf{n}) = 1/q(\mathbf{n})$
	
	\BlankLine
	
	\KwResult{Value of the h-convexified function $\Gamma$ at $\mathbf{B}$:  $\Gamma(\mathbf{B})=\gamma(\mathbf{n})$}
	
	\BlankLine
	\BlankLine
	\BlankLine
	\BlankLine
	
	\caption{Sketch of the subroutine for determining h-convex $\Gamma$ for arbitrary 
		boundary $\mathbf{B}$ based on energies $\Gamma_0^{\kappa}$
		for a discrete set of boundaries $\mathbf{B}^{\kappa}$.}

\end{algorithm}

\begin{figure}
	\begin{picture}(300,280)(0,0)
		\put(75,10){\resizebox{4.5 cm}{!}{\includegraphics{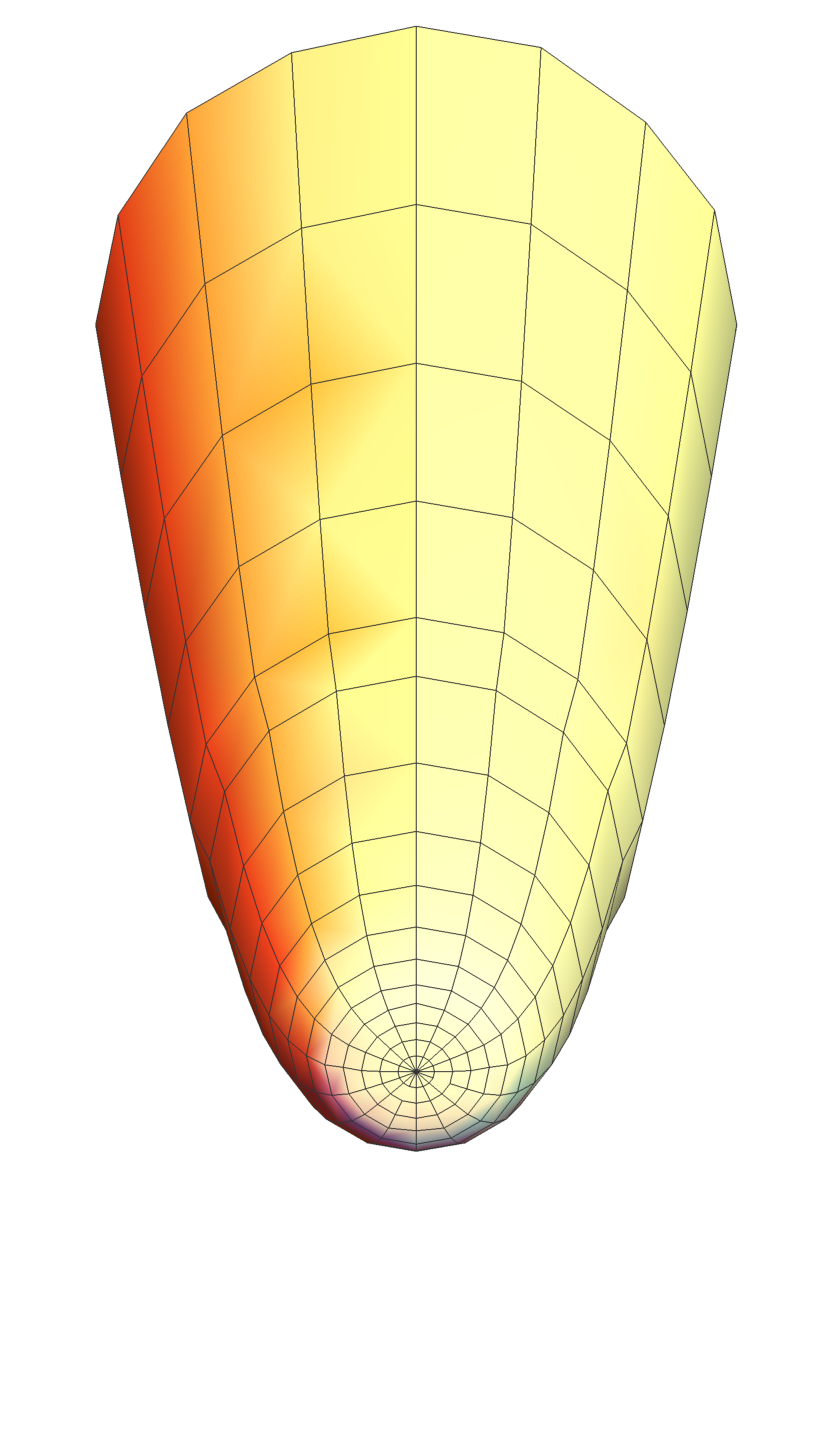}}}
		\put(230,0){\resizebox{4.5 cm}{!}{\includegraphics{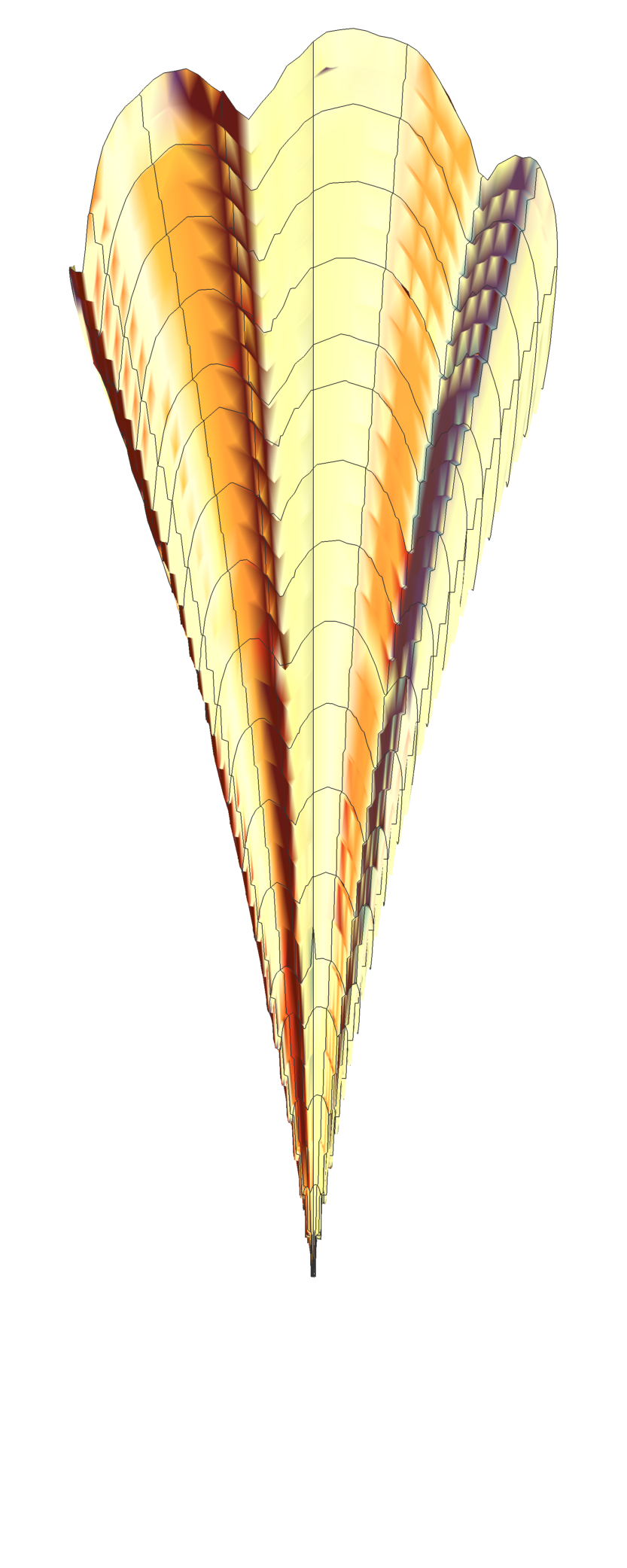}}}		
		\put(75,180){\textit{a}}
		\put(230,270){\textit{b}}
		\put(173,80){$\gamma_0$ }
		\put(305,80){$\gamma$ }
	\end{picture}
	\vskip 0.0cm
	\caption{The dips in $\gamma$-plots at $[1 \, 1 \,1]$ for $\Sigma 3$ misorientation 
		modeled by the interpolating function $\gamma_0$ 
		obtained by inverse distance weighting 
		of data 
		from \cite{olmsted2009survey} (\textit{a}) 
		and $\gamma$ obtained by h-convexification of $\gamma_0$ (\textit{b}). 
		The maximal deviation of shown directions from $[1 \, 1 \,1]$ is $3^{\circ}$. 
		The function $\gamma_0$ is differentiable
		whereas $\gamma$ lacks differentiability at $[1 \, 1 \,1]$
		(i.e., at the projectile point of the `arrowhead')
		and along edges of the `arrowhead'.
	}
	\label{Fig_cusp_shape}
\end{figure}
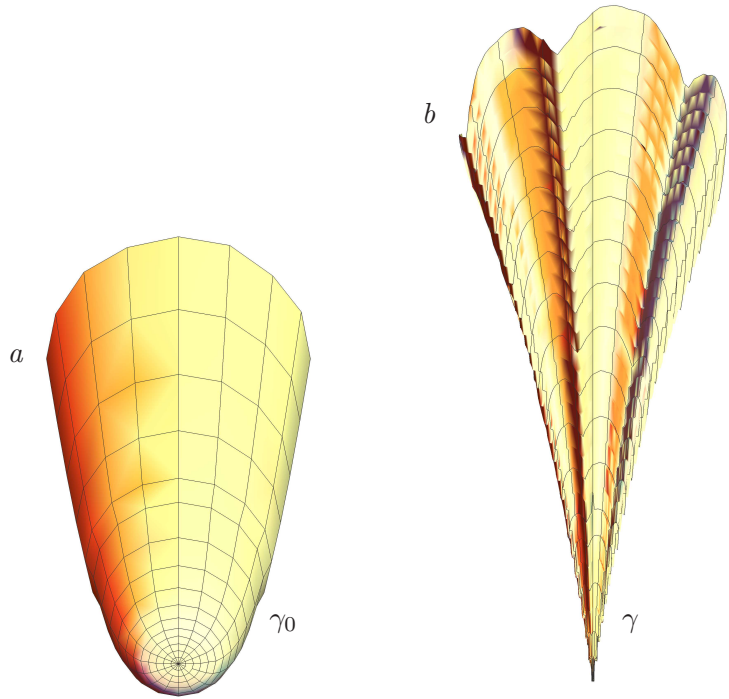

\subsection{Example}

\begin{figure}
	\begin{picture}(300,580)(0,0)
		\put(30,400){\resizebox{5.7 cm}{!}{\includegraphics{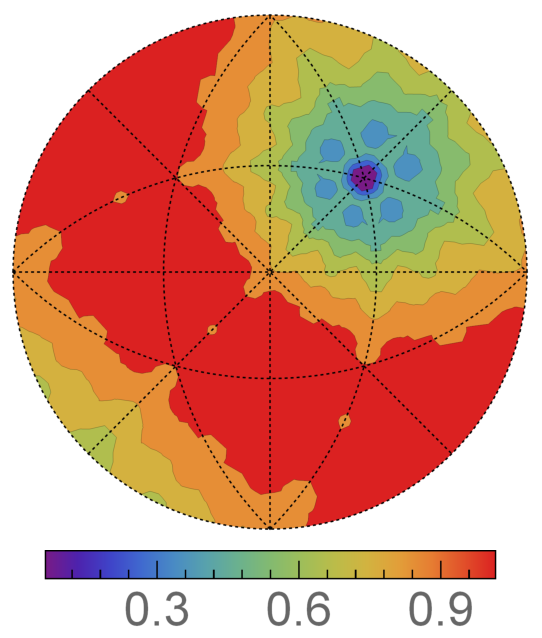}}}
		\put(30,200){\resizebox{5.7 cm}{!}{\includegraphics{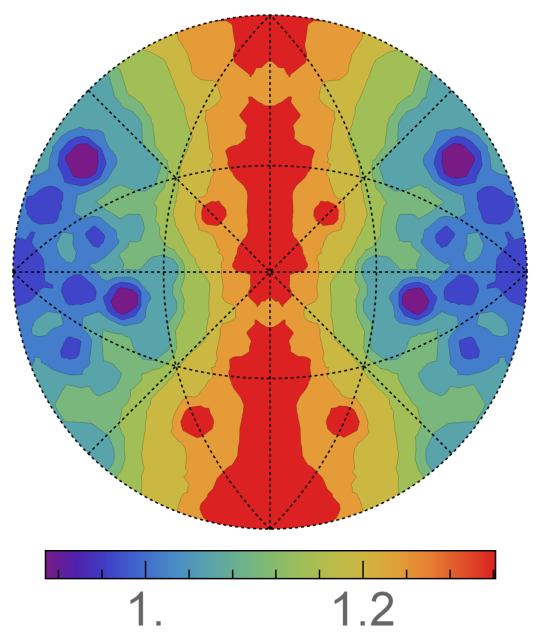}}}
		\put(30,0){\resizebox{5.7 cm}{!}{\includegraphics{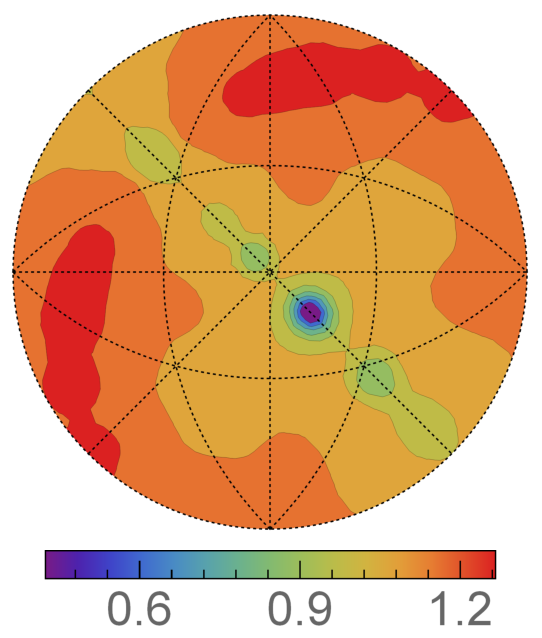}}}		
		\put(240,400){\resizebox{5.7 cm}{!}{\includegraphics{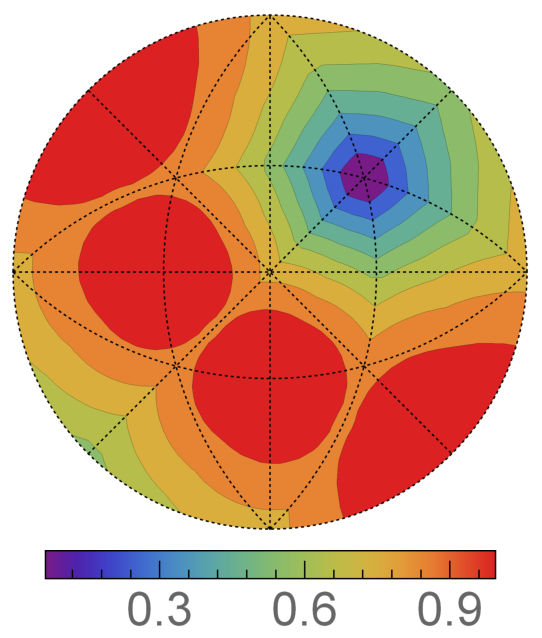}}}
		\put(240,200){\resizebox{5.7 cm}{!}{\includegraphics{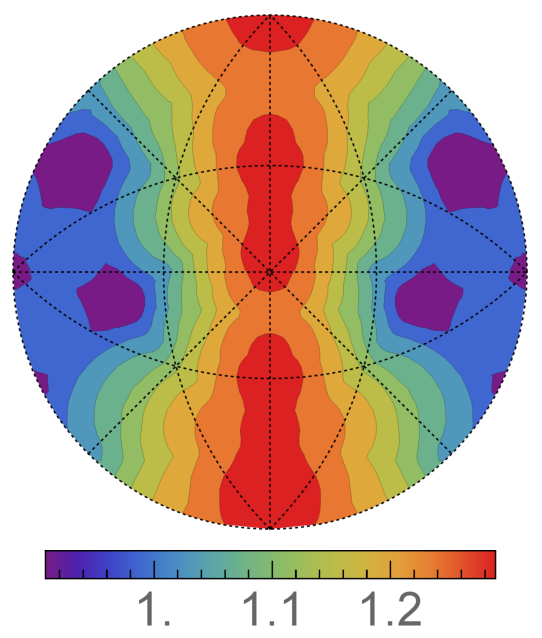}}}
		\put(240,0){\resizebox{5.7 cm}{!}{\includegraphics{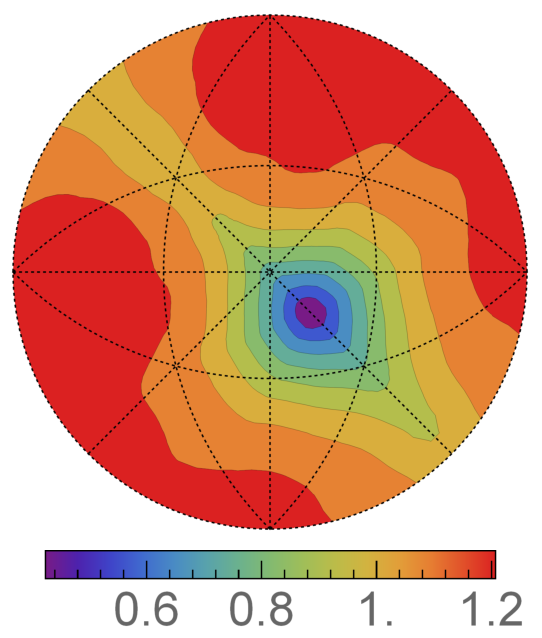}}}		
		\put(20,585){\textit{a}}
		\put(20,385){\textit{b}}
		\put(20,185){\textit{c}}
		\put(160,585){\large $\Gamma_0$}
		\put(370,585){\large $\Gamma$}	
	\end{picture}
	\vskip 0.0cm
	\caption{Energy represented by the interpolating function $\Gamma_0$ 
		(left column)
		and h-convex function $\Gamma$ (right column)
		for the misorientations  
		$\Sigma 3$  (\textit{a}),
		$\Sigma 5$  (\textit{b}) and
		$\Sigma 11$  (\textit{c}). 
		The energy unit is $\mbox{J/m}^2$. 
	}
	\label{Fig_resulting_energy_function}
\end{figure}
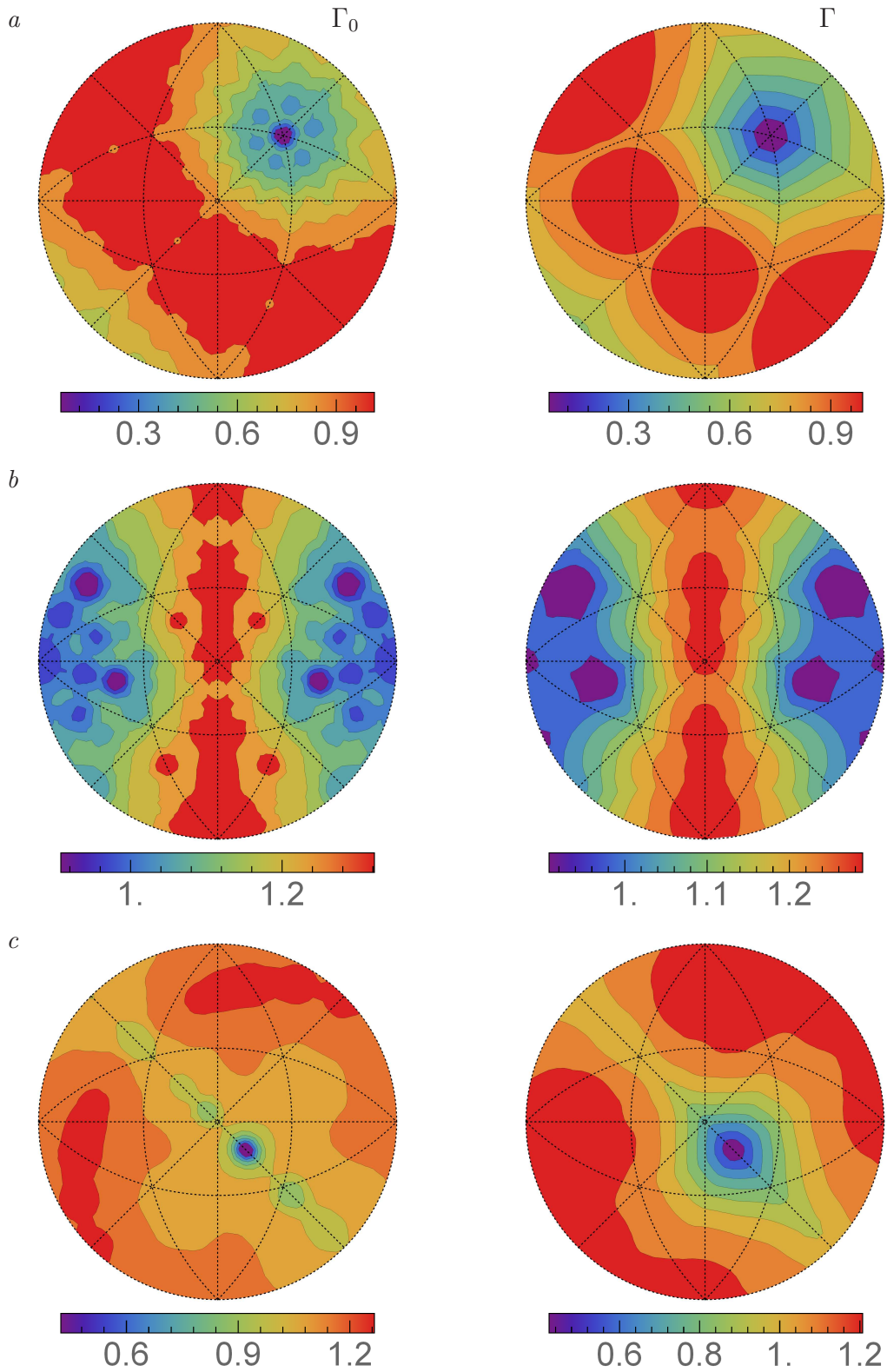

H-convexification will be illustrated on 
data of Olmsted et al. \cite{olmsted2009survey}.
The dataset contains energies $\Gamma_0^{\kappa}$ for $K=388$ boundary types
$\mathbf{B}^{\kappa}$ ($\kappa=1,2,\ldots,K$).
The set was appended with the point of zero energy at $\mathbf{B}^0$.
The preliminary function $\Gamma_0$ was constructed by interpolation. 
A simple method for interpolating between irregularly spaced data points in metric spaces 
is Shepard's inverse distance weighting \cite{Shepard_1968}.
Let $\widetilde{d}$ be a symmetry-invariant distance 
function on the boundary space \cite{Morawiec_2019_AM}.
To get 
the value of $\Gamma_0$ at $\mathbf{B}$
one needs 
the distances $d_{\kappa}= \widetilde{d}(\mathbf{B},\mathbf{B}^{\kappa})$ 
of $\mathbf{B}$ to data points $\mathbf{B}^{\kappa}$.
The interpolated value $\Gamma_0(\mathbf{B})$ is the arithmetic
average of $\Gamma_0^{\kappa}$ weighted by $d_{\kappa}^{-u}$ $(u > 0)$.
Unless stated otherwise, the parameter $u$ was set to 4,  
and boundary distances were computed using the function $\widetilde{D}_M$ defined in  \cite{Morawiec_1998}.\footnote{Since 
	$\mathbf{B}^0$ has 
	multiple representations $(I,\mathbf{n})$,
	where $I$ is the identity matrix 
	and $\mathbf{n}$ is an arbitrary unit vector, 
	the distance from $\mathbf{B}$ to $\mathbf{B}^0$ 
	is $d_0 = \min_{\mathbf{n}} \widetilde{d}(\mathbf{B},(I,\mathbf{n}))$. 
	For  distance functions denoted in \cite{Morawiec_2019_AM} by a capital letter $D$, 
	$d_0$ is the misorientation angle of $\mathbf{B}$
	times a constant factor.}
The interpolating function $\Gamma_0$ calculated in this way 
is not only continuous but also  differentiable\footnote{With $u>1$, first derivatives 
of $\Gamma_0$ at data points $\mathbf{B}=\mathbf{B}^{\kappa}$ are zero.}
(Fig.~\ref{Fig_cusp_shape}\textit{a}). 
The function $\Gamma_0$ obtained from data of Olmsted et al. \cite{olmsted2009survey} was not h-convex.
It was h-convexified using $N_P \approx 2.4 \times 10^3$ auxiliary directions.
If the direction $\mathbf{n}_p$ was used, 
so were the directions symmetrically equivalent to it.
Convex hulls were calculated using the code 
of O'Rourke \cite{oRourke1998code,oRourke1998}.
Example sections through the interpolating function $\Gamma_0$ 
and the function $\Gamma$ resulting from h-convexification of $\Gamma_0$
are shown in Fig.~\ref{Fig_resulting_energy_function}.
Since the computation of $\Gamma$ is based on a finite number 
of points, individual $1/\gamma$-plots are polyhedra, and in effect,
$\Gamma$ has multiple singularities at constant misorientations.
Due to the high value of $N_P$, most of them are unnoticeable, but some are clearly 
visible (Fig.~\ref{Fig_cusp_shape}\textit{b}).

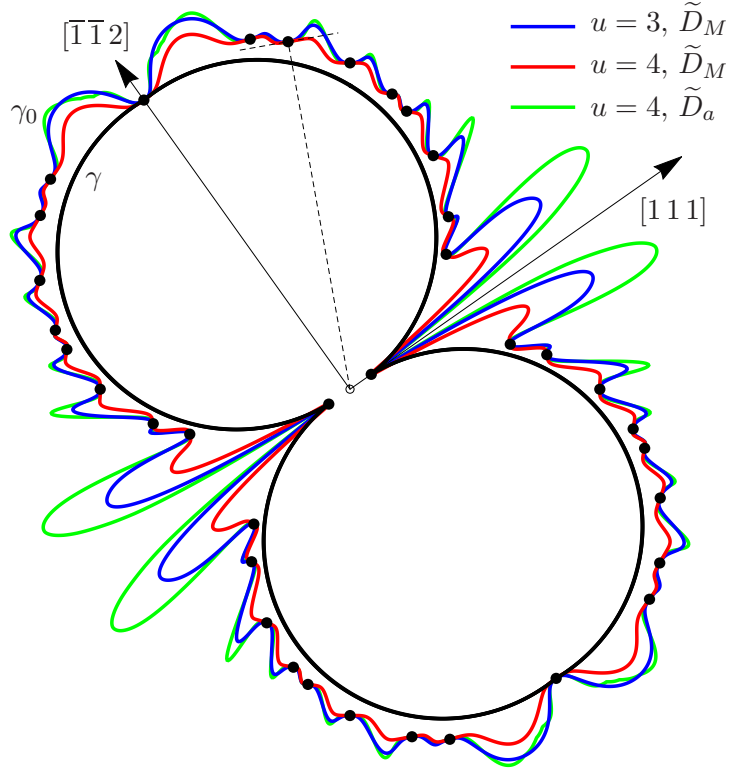
\begin{figure}
	\begin{picture}(300,300)(0,0)
		\put(80,0){\resizebox{9.0 cm}{!}{\includegraphics{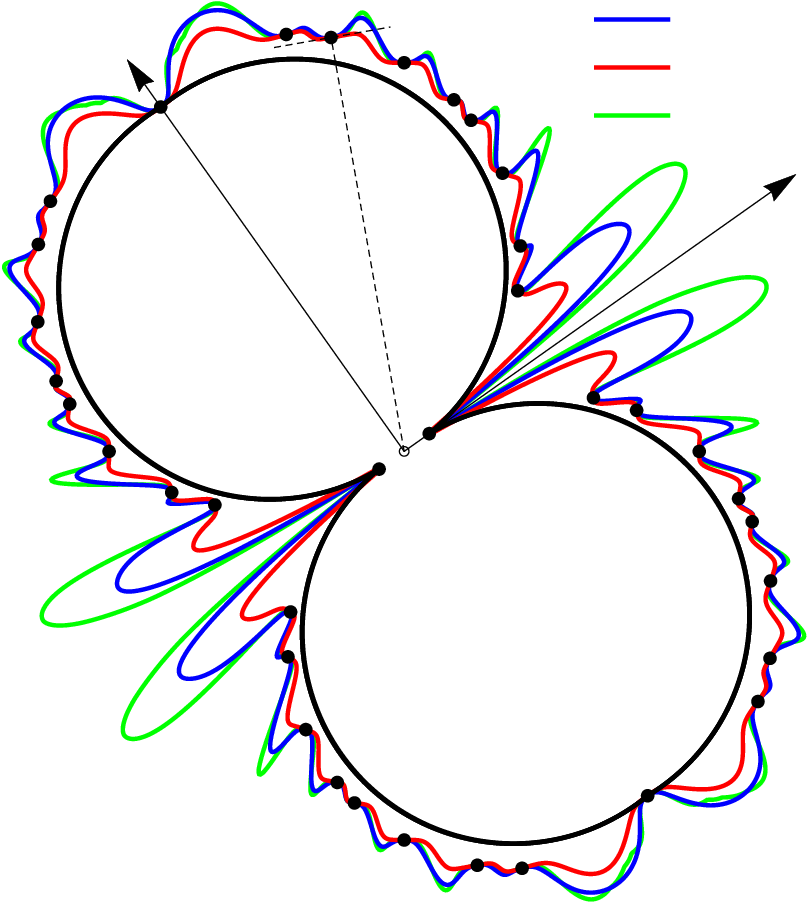}}}
		\put(81,245){$\gamma_0$}
		\put(109,220){$\gamma$}
		\put(100,273){$[\overline{1} \, \overline{1} \, 2 ]$}
		\put(316,207){$[1 \, 1 \, 1 ]$}
		\put(298,278){$u=3$, $\widetilde{D}_M$}
		\put(298,262){$u=4$, $\widetilde{D}_M$}
		\put(298,246){$u=4$, $\widetilde{D}_a$}	
	\end{picture}
	\vskip 0.0cm
	\caption{Section perpendicular to 
		$[1 \, \overline{1} \, 0]$ through Wulff plots 
		of the interpolating functions $\gamma_0$ and h-convex $\gamma$ 
		for $\Sigma 3$.  
		The plots are for $u=3$ and $u=4$ and 
		for the distances $\widetilde{D}_a$ and $\widetilde{D}_M$.
		H-convexification of functions $\gamma_0$ 
		leads to nearly identical resulting functions
		$\gamma$; they are all shown in black. 
		Disks mark data points from \cite{olmsted2009survey}. 
		The magnitude of vectors is $1\,\mbox{J/m}^2$.
	}
	\label{Fig_mod_S3_1m10}
\end{figure}

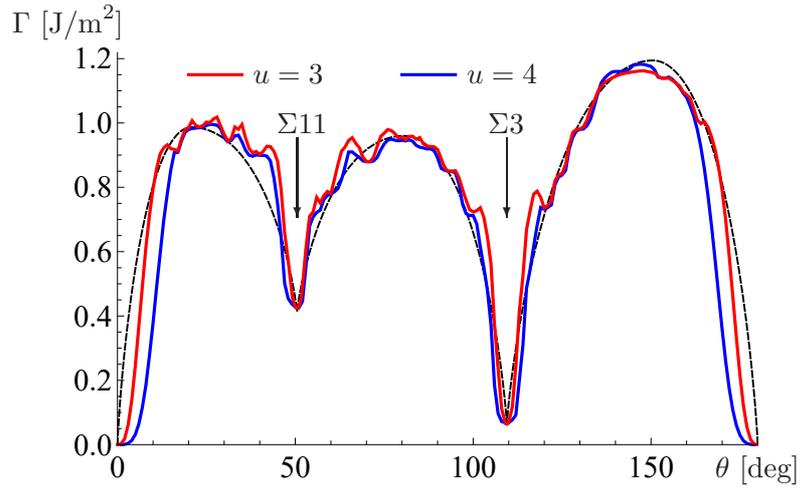
\begin{figure}
	\begin{picture}(300,180)(0,0)
		\put(80,0){\resizebox{9.0 cm}{!}{\includegraphics{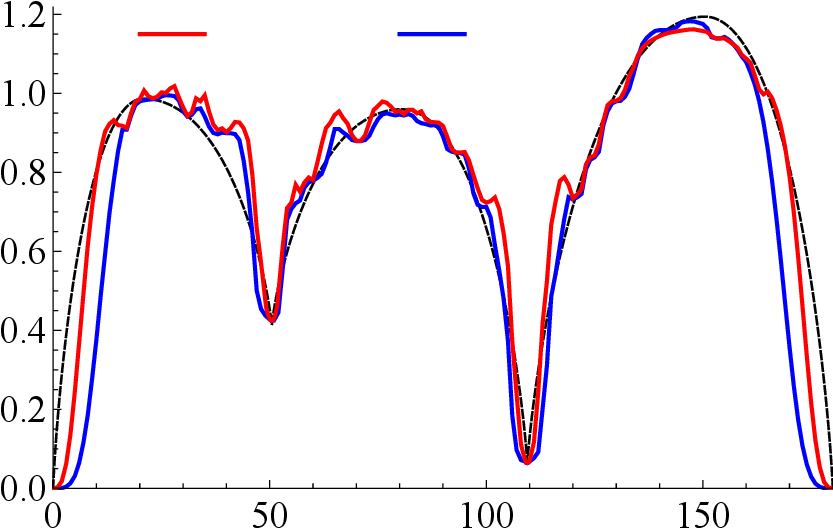}}}
		\put(57,168){$\Gamma$ [$\mbox{J/m}^2$]}
		\put(320,0){$\theta$ [deg]}
		\put(147,149){$u=3$}
		\put(227,149){$u=4$}
		\put(156,130){$\Sigma 11$}
		\put(163.5,128){\vector(0,-1){30}}
		\put(235,130){$\Sigma 3$}
		\put(242,128){\vector(0,-1){30}}
	\end{picture}
	\vskip 0.0cm
	\caption{H-convex $\Gamma$ function for $[ 110 ]$ symmetric tilt boundaries
		versus the tilt angle $\theta$. 
		The interpolating functions $\Gamma_0$ (not shown) were computed with $u=3$ 
		(red graph) and $u=4$ (blue graph).
	Dashed black line represents the function described in \cite{bulatov2014grain}.
	}
	\label{Fig_Tilt_110}
\end{figure}

Clearly, the shape of the $\Gamma$ function 
is influenced by the choice of the interpolation method, 
but this effect is mitigated by h-convexification.
This can be observed by testing $\Gamma_0$ 
and $\Gamma$ for different values the $u$ parameter and different distance functions.
When $u$ is small, the interpolating function is strongly attracted to 
the arithmetic mean $\sum_{\kappa} \Gamma_0^{\kappa}/K$
and exhibits large oscillations.
When $u$ is large, results of 
inverse distance weighting are close to those of
the nearest-neighbor interpolation and this dulls extrema.
Example functions for $u=3$, $u=4$ 
and for the distance functions $\widetilde{D}_M$ \cite{Morawiec_1998} and $\widetilde{D}_a$ \cite{Morawiec_2019_MMT}
are compared in Fig.~\ref{Fig_mod_S3_1m10}.
This figure also shows that the 
character of the function $\Gamma$ resulting from 
h-convexification can differ considerably 
from that of the preliminary $\Gamma_0$.
Sections through h-convex $\Gamma$ for fixed misorientation are simple
compared to those through the interpolating function $\Gamma_0$, 
and edges and vertices of convex hulls of $1/\gamma$-plots 
give rise to singularities with respect to boundary inclination parameters. 
This, however, does not apply to misorientation parameters. 
The dependence of $\Gamma$ on misorientation, 
with unjustified oscillations and unphysical behavior near zero-misorientation,
reflects the underlying interpolating function;
see Fig.~\ref{Fig_Tilt_110}.
The well known deficiencies of the interpolation by inverse distance weighting \cite{Shepard_1968},
which are largely eliminated from inclination-dependent sections through $\Gamma$,
remain visible in misorientation-dependent functions.
They could be removed by using a more sophisticated interpolation method.
However,  the space of macroscopic boundary parameters 
has high dimension and nontrivial geometry,
and one needs to take into account that distributions of data points 
in the space of macroscopic boundary parameters are usually highly non-uniform.
Under these conditions, devising a simple but effective interpolation method is challenging.

	The resulting h-convex energy function $\Gamma$ is similar but not the same as that 
	of Bulatov et al. \cite{bulatov2014grain}; 
	cf. right columns of Figs.~\ref{Fig_example_Wulff_plots} 
	and \ref{Fig_resulting_energy_function}.
	These two functions are based on different principles. 
	First,  the function $\Gamma$ is computed without assumptions 
	about locations of energy cusps,
	whereas that of Bulatov et al. \cite{bulatov2014grain}  
	is based on presupposed locations of cusps.
	Second, $\Gamma$ is computed without assumptions about profiles of the cusps, 
	whereas  Bulatov et al. \cite{bulatov2014grain} use the RSW function. 
	Third, the number of parameters  governing determination of $\Gamma$ is small  
	($u$ and $N_P$ plus the choice of the distance function),
	whereas some forty parameters are used in \cite{bulatov2014grain}. 
	Finally, $\Gamma$ satisfies Herring's condition,
	whereas the function described in \cite{bulatov2014grain}  
	does not (but could serve as $\Gamma_0$). 
	On the other hand, due to the use of a simplistic interpolation method, 
	the function $\Gamma$ shows unphysical features in the 
	energy-misorientation dependence.

	As for the computational cost of the described procedure,
	by far the most expensive part of acquiring data for
	plots like those in Fig.~\ref{Fig_resulting_energy_function} 
	is the interpolation. 
	It takes over 99\% of the total computation time.
	The cost of interpolation is proportional to $N_P K N_s^2$, 
	where $N_s$ is the number of used crystal symmetry operations. 
	In the considered example (with $N_s=24$), 
	sequential processing on a 2.6 GHz personal computer
	took about six minutes per misorientation. 
	In the case of Fig.~\ref{Fig_Tilt_110}, 
	each point on the abscissa corresponds to a different misorientation. 
	Therefore, obtaining a single point of the one-dimensional color plot
	in Fig.~\ref{Fig_Tilt_110} takes nearly the same time as obtaining
	all points for one of the two-dimensional plots
	in Fig.~\ref{Fig_resulting_energy_function}.

\section{Concluding remarks}

The paper addresses the issue of 
proper modeling of 
grain boundary energy as
a function of macroscopic boundary parameters.
Energies represented by the function are assumed to correspond 
to boundary equilibrium states.
For arbitrary constant misorientation, 
the function should 
satisfy the condition that 
the $1/\gamma$-plot is convex.
This property of a global boundary energy function is referred to 
as h-convexity.
It is pointed out that  
current accounts on determination of boundary energy functions 
ignore the requirement of h-convexity. 
Moreover, it is shown that 
if energy cusps at fixed misorientation are modeled using 
Read-Shockley-Wolf function, 
the h-convexity condition is violated.

A function that approximates the true energy function 
but violates the h-convexity condition 
can be `h-convexified'
by simply replacing all high function values 
causing the violation 
by smaller (but the largest possible) values 
with which the condition is satisfied.  
The function resulting from the h-convexification may have 
features expected for an energy function, 
even if the input function lacks them.
In particular, the h-convexification may lead 
to the appearance of singularities not present in the input function.

The paper provides details of an
example procedure for constructing continuous 
h-convex function based on a discrete dataset of energies. 
The emphasis is on simplicity.
The procedure comes down to h-convexification
of a function obtained by
Shepard's interpolation 
and is governed by a small number of parameters.
It does not involve a priori selected  locations of singularities
and does not use any extrinsic cusp models. 
Application of the procedure to simulated data of Olmsted et al. \cite{olmsted2009survey} 
leads to a global function similar to that 
of Bulatov et al. \cite{bulatov2014grain}, 
but it requires much fewer assumptions and parameters.
Clearly, the nature of the data source is irrelevant, 
and the procedure can be applied to other types of data, 
e.g., experimental results or 
combinations of data from different sources.

The described approach can be seen as a step 
towards more advanced strategies for modeling of grain boundary energy.
Various improvements can be envisioned,   
but the key issue is to develop a better interpolation method.
With an appropriate interpolation scheme, 
energy singularities in the domain of misorientations (other than that at zero misorientation)
could be linked to singularities arising in the domain of plane inclinations
(due to vertices and edges of $1/\gamma$-plots).
A properly chosen modeling strategy would allow for avoiding assumptions 
about locations of singularities.

One might also consider adding the condition that would prevent 
a configuration to achieve lower energy by replacing 
a single grain boundary with two or more boundaries parallel to it.	
In other words, energy of a boundary needs to be lower than the sum energies 
of any possible replacements. 
If the energy distribution is close to uniform,
the condition is satisfied, but theoretically, 
a general energy function may violate it.

\clearpage

\bibliographystyle{unsrt}
\bibliography{GBEnergy.bib}

\end{document}